\documentclass[aps,pra,twocolumn,showpacs,groupedaddress,amsmath,amssymb]{revtex4}

\usepackage{graphicx}

\usepackage{color}
\usepackage{bm}
\usepackage{latexsym}

\begin{document}

\title {Exceptional Point Dynamics in Photonic Honeycomb Lattices with ${\cal PT}$ Symmetry}
\author{Hamidreza Ramezani}
\author{Tsampikos Kottos}
\affiliation{Department of Physics, Wesleyan University, Middletown, Connecticut 06459, USA}
\affiliation{MPI for Dynamics and Self-Organization - Bunsenstra\ss e 10, D-37073 G\"{o}ttingen, Germany}

\author{Vassilios Kovanis}
\affiliation{Air Force Research Laboratory, Sensors Directorate, Wright Patterson AFB, OH 45433 USA}

\author{Demetrios N. Christodoulides}
\affiliation{College of Optics \& Photonics-CREOL, University of Central Florida, Orlando, Florida 32816, USA}

\date{\today}
\begin{abstract}

{We theoretically investigate the flow of electromagnetic waves  in complex honeycomb photonic lattices with local 
${\cal PT}$ symmetries.   Such  ${\cal PT}$ structure is  introduced via a judicious arrangement of gain or loss 
across the honeycomb lattice, characterized by a gain/loss  parameter $\gamma$. We found a new class of conical 
diffraction  phenomena where the formed cone is brighter and travels along the lattice with a transverse speed 
proportional to $\sqrt{\gamma}$.}
\end{abstract}

\maketitle
\section{Introduction}

{\it Conical refraction}  phenomena i.e the spreading into a hollow cone of an unpolarized light beam entering a 
biaxial crystal along its optic axis, is fundamental in classical optics and in mathematical physics \cite{BJ07,
IN06,H37,L37}. Originally predicted by Hamilton in 1837 \cite{H37} and  experimentally observed by Lloyd \cite{L37}, 
these phenomena have been intensively studied in recent years, by a large community of theorists and experimentalists 
\cite{BJ07,IN06,P07,T08,ANZ09,H37,L37,E06}. The physical origin of the phenomenon is associated  with the existence 
of the legendary {\it diabolical points}, that emerge along the axis of intersection of the two shells associated 
with the wave surface. Around a diabolical point the energy dispersion relation is linear while the direction of 
the group velocity is not uniquely defined. Recently conical diffraction was observed in $2D$ photonic honeycomb 
lattices \cite{P07} which shares key common features, including the existence of diabolical points, with the band 
structure of graphene in condensed matter physics literature. In graphene, the electrons around the diabolical points 
of the band structure behave as  {\it massless relativistic fermions}, thus resulting in extremely high electron 
mobility. Both photonic and electronic graphene structures allow us to test experimentally various legendary predictions 
of relativistic quantum mechanics such  as the Klein paradox \cite{BPGSSP10}, and the dynamics of optical tachyons 
\cite{SRBS11}.

While diabolical points  are spectral singularities associated with Hermitian systems, for pseudo-Hermitian Hamiltonians, 
like those used for the theoretical description of non-Hermitian optics, a topologically different singularity may appear: 
an {\it exceptional point} (EP), where not only the eigenvalues but also the associated eigenstates coalesce. Pseudo-Hermitian 
optics is a rapidly developed field which aims via a judicious design that involves the combination of delicately balanced 
amplification and absorption regions together with the modulation of the index of refraction, to achieve new classes of 
synthetic meta-materials that can give rise to altogether new physical behavior and novel functionality \cite{MGCM08a,K10}. 
The idea can be carried out via index-guided geometries with special antilinear symmetries. Adopting a Schr\"odinger language 
that is applicable in the paraxial approximation, the effective Hamiltonian that governs the optical beam evolution is non
-Hermitian and commutes with the combined parity (${\cal P}$) and time (${\cal T}$) operator \cite{BB98,BBDJ02,B07}. In 
optics, ${\cal PT}$ symmetry demands that the complex refractive index obeys the condition $n({\mathbf r})=n^*(-{\mathbf r})$. 
It can be shown that for such structures, a real propagation constant (eigenenergies in the Hamiltonian language) exists for 
some range (the so-called {\it exact phase}) of the gain/loss coefficient. For larger values of this coefficient the system 
undergoes a {\it spontaneous symmetry breaking}, corresponding to a transition from real to complex spectra (the so-called 
{\it broken phase}). The phase transition point, shows all the characteristics of an {\it exceptional point} (EP) singularity. 

${\cal PT}$ symmetries are not only novel mathematical curiosities. In a series of recent experimental papers ${\cal PT}$ 
dynamics have been investigated and key predictions have been confirmed and demonstrated \cite{RMGCSK10,GSDMRASC09,
SLZEK11,FAHXLCFS11}. Symmetry breaking has been experimentally observed in non-Hermitian structures \cite{RMGCSK10,
GSDMRASC09,SLZEK11} while power law growth -- characteristic of phase transitions -- of the total energy has been 
demonstrated close to the {\it exceptional points} in Ref.~\cite{SLZEK11}. In a silicon platform claims have been made that 
non-reciprocal light propagation in a silicon photonic circuit has been recorded \cite{FAHXLCFS11}. ${\cal PT}$-synthetic 
materials can exhibit several intriguing features. These include among others, power oscillations and non-reciprocity 
of light propagation \cite{MGCM08a,RMGCSK10,ZCFK10,JG11}, non-reciprocal Bloch oscillations \cite{L09a}, and unidirectional 
invisibility \cite{LRKCC11}. More specifically, a recent paper has proposed photonic honeycomb lattices with ${\cal 
PT}$ symmetry \cite{SRBS11}. Interestingly, that work has shown that introducing alternating gain/loss to a honeycomb 
system prohibits the $\cal{PT}$-symmetry, but adding appropriate strain (direction and strength) restores the symmetry, 
giving rise to $\cal{PT}$-symmetric photonic lattices \cite{SRBS11}. Moreover, the mentioned work has found that in 
such systems, much higher group velocities can be
achieved (compared with non $\cal{PT}$-symmetry breaking systems), corresponding to a tachyonic dispersion relation \cite{SRBS11}. 
In the nonlinear domain, such pseudo-Hermitian non-reciprocal effects can be used to realize a new generation of 
on-chip isolators and circulators \cite{RKGC10}. Other results within the framework of ${\cal PT}$-optics include 
the realization of coherent perfect laser absorber \cite{L10b}, spatial optical switches \cite{FMM}, and nonlinear switching structures \cite{SXK10}. 
Despite the wealth of results on transport properties of ${\cal PT}$-symmetric $1D$ optical structures, the 
properties of high dimensional ${\cal PT}$ optical lattices, (with the exception of few recent studies 
\cite{MGCM08a,Mussli}), has remained so far essentially unexplored.

Recently, it was pointed \cite{M02a} that ${\cal PT}$-symmetric Hamiltonians are a special case
of pseudo-Hermitian Hamiltonians i.e. Hamiltonians that have an {\it antilinear} symmetry \cite{M02b,M08,BBM02}.
Such Hamiltonians commute with an antilinear operator ${\cal ST}$, where ${\cal S}$ is a generic linear
operator. The corresponding Hamiltonian ${\cal H}$ is termed {\it generalized} ${\cal PT}$-symmetric. In a
similar manner as in the case of ${\cal PT}$-symmetry, one finds that if the eigenstates of ${\cal H}$ are
also eigenstates of the ${\cal ST}$-operator then all the eigenvalues of ${\cal H}$ are strictly real and 
the ${\cal ST}$-symmetry is said to be {\it exact}. Otherwise the symmetry is said to be {\it broken}.
An example case of generalized ${\cal PT}$-symmetric optical structure is the one reported in \cite{BFKS10}
(see also \cite{ZCFK10}).
The unifying feature of these systems is that they are built of a particular kind of ``building blocks'' which
below are referred to as {\it dimers}. Each dimer in itself does have ${\cal PT}$-symmetry and it
can be represented as a pair of sites with assigned energies $\epsilon_{n},\epsilon_{n}^{\ast}$. The system 
is composed of such dimers, coupled in some way. For an arbitrary choice of
the site energies $\epsilon_{n}$ and coupling between the dimers, the system as a whole does not possess ${\cal PT}$-symmetry (indeed,
such  {\it global} ${\cal PT}$-symmetry would require precise relation between various $\epsilon_{n}$ and coupling
symmetry between the dimers).
On the other hand, since each dimer is ${\cal PT}$-symmetric, with respect to its own center, there
is  some kind of ``local'' ${\cal PT}$-symmetry (which we shall define as ${\cal P}_d{\cal T}$-symmetry).
The main message of these papers is that, ${\cal P}_d{\cal T}$-symmetry ensures a robust region of parameters
in which the system has an entirely real spectrum. Below, we will use the term ${\cal PT}$-symmetry in a loose
manner and we will include also systems with generalized ${\cal PT}$-symmetry. 

In this paper we investigate beam propagation in non-Hermitian two dimensional  photonic honeycomb lattices with 
${\cal PT}$-symmetry and probe for the possibility of abnormal diffraction. We find a new type of conical diffraction 
that is associated with the spontaneously ${\cal PT}$-symmetry breaking phase transition point. Despite the fact 
that at the EP the Hilbert space collapses the emerging cone is brighter and propagates with a transverse velocity 
that is controlled by the gain and loss parameter $\gamma$.

The organization of the paper is as follows: In Section \ref{PLM}, we present the mathematical model. In Section 
\ref{eimodes} we analyze the stationary properties of the system and introduce a criterion based on the degree of 
non-orthogonality of the eigenmodes in order to identify the EP for finite system sizes. The beam dynamics is studied 
in Section \ref{dynamics}. First, we study numerically the beam evolution in subsection \ref{numerics} while in
the theoretical analysis is performed in subsection \ref{theory}. Our conclusions are given at the last Section \ref{conclusions}.

\begin{figure}
   \includegraphics[width=.75\linewidth, angle=0]{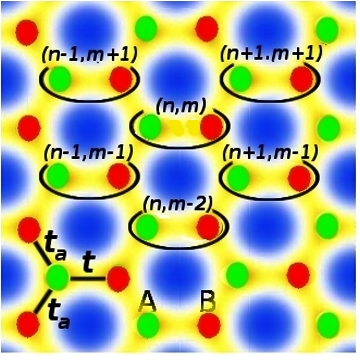}
    \caption{Honeycomb photonic lattice structure with intra-dimer coupling ${t}$ and inter-dimer coupling ${t_a=1}$. Sub-
lattice (lossy waveguide) $a_{n,m}$ is shown by green circles while sub-lattice (gain waveguide) $b_{n,m}$ is shown 
by the red circles. Each dimer is distinguished by index ${n}$ and ${m}$. The field is coupled evanescently between
the waveguides.
\label{fig:figure1}}
\end{figure}

\section{Photonic Lattice Model}\label{PLM}

We consider a two dimensional honeycomb photonic lattice of coupled optical waveguides. Each 
waveguide supports only one mode, while light is transferred from waveguide to waveguide through optical 
tunneling. A schematic of the set-up is shown in Fig. \ref{fig:figure1}. The lattice consist of two types of waveguides: 
type (A) made from lossy material (green) whereas type (B) exhibits the equal amount of gain (red). Their arrangement in space is 
such that they form coupled (A-B) dimers with inter and intra-dimer couplings $t_a$ and $t$ respectively. Such structure,
apart from a global ${\cal PT}$ symmetry, respects also another anti-linear symmetry (in Ref. \cite{BFKS10} we 
coined this ${\cal P}_d{\cal T}$-symmetry) which is related with the local ${\cal PT}$-symmetry of each individual dimer.

Without loss of generality we rescale everything in units of the inter-dimer coupling $t_a=1$. In the tight binding 
description, the diffraction dynamics of the mode electric field amplitude $\Psi_{n,m}=(a_{n,m},b_{n,m})^T$ at the 
$(n,m)-$th dimer evolves according to the following Schr\"odinger-like equation
\begin{equation}
\label{dyndimer}
\begin{array}{lcr}
i {da_{n,m}\over dz} + \epsilon a_{n,m} + b_{n-1,m+1} + b_{n-1,m-1} + t b_{n,m} &=&0\\
i {db_{n,m}\over dz} + \epsilon^* b_{n,m} + a_{n+1,m+1} + a_{n+1,m-1} + t a_{n,m}&=&0
\end{array}
\end{equation}
where $\epsilon=\epsilon_0+i\gamma$ is related to the refractive index \cite{MGCM08a}. Without loss of generality, 
we assume $\epsilon_0=0$ and $\gamma>0$. 

For latter purposes, it is useful to work in the momentum space. To this end, we write the field amplitudes $\Psi_{n,m}$ 
in their Fourier representation i.e.
\begin{equation}
\label{four}
\begin{array}{lcr}
 a_{n,m}(z)&=&{1\over (2\pi)^2}\int_{-\pi}^{\pi} dk_x dk_y {\tilde a}_{k_x,k_y}(z)\exp(i [n k_x+m k_y]) \\
 b_{n,m}(z)&=&{1\over (2\pi)^2}\int_{-\pi}^{\pi} dk_x dk_y {\tilde b}_{k_x,k_y}(z)\exp(i [n k_x+m k_y]).
\end{array}
\end{equation}
Substitution of Eq.~(\ref{four}) to Eq.~(\ref{dyndimer}) leads to the following set of coupled differential equations
in the momentum space
\begin{equation}
\label{dynfourier}
i{d\over dz}\left(
\begin{array}{c}
{\tilde a}_{\mathbf k}(z)\\
{\tilde b}_{\mathbf k}(z)
\end{array}
\right)
= H_{\mathbf k}
\left(
\begin{array}{c}
{\tilde a}_{\mathbf k}(z) \\
{\tilde b}_{\mathbf k}(z)
\end{array}
\right)
\end{equation}
where 
\begin{equation}
\label{hamilt}
 \quad
H_{\mathbf k}=\left(
\begin{array}{cc}
 -i\gamma & D({\mathbf k})\\
 D({\mathbf k})^* & i\gamma
\end{array}
\right).
\end{equation}
and
\begin{equation}
 D(\mathbf k)=-(t+2e^{-i k_x}\cos k_y);{\mathbf k} \equiv (k_x, k_y)
\end{equation}
In other words, because of the translational invariance of the system, the equations of motion in the
Fourier representation break up into $2\times 2$ blocks, one for each value of momentum ${\bf k}$. The 
two-component wave functions for different ${\bf k}$ values are then decoupled, thus allowing for a simple
theoretical description of the system. 

\begin{figure}
   \includegraphics[width= 1\linewidth, angle=0]{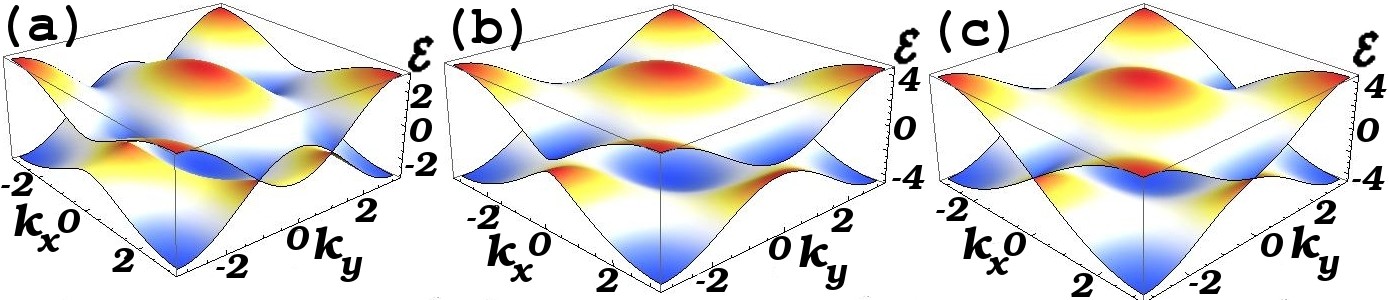}
    \caption{Energy surfaces for a honeycomb lattice with $t_a=1$ and (a) $(t,\gamma)=(1,0)$ where a DP is present; 
(b) $(t,\gamma)=(2.5,0)$ where a band-gap has been created and (c) $(t,\gamma)=(2.5 ,0.5)$ where a gain/loss parameter
led to the creation of a EP.
\label{figure:fig2}}
\end{figure}

{\section{Eigenmodes Analysis}\label{eimodes}

We start our analysis with the study of the stationary solutions corresponding to the Eq.~(\ref{dynfourier}). 
Substituting the stationary form 
\begin{equation}
 ({\tilde a}_{n,m},{\tilde b}_{n,m})^T= \exp(-i {\cal E}z)(A,B)^T
\end{equation}
in Eq.~(\ref{dynfourier}) we get
\begin{equation}
\label{stationary}
{\cal E}\left(
\begin{array}{c}
A\\
B
\end{array}
\right)
= H_{\mathbf k} 
\left(
\begin{array}{c}
A\\
B
\end{array}
\right)
\end{equation}
The spectrum is obtained by requesting a non-trivial solution i.e. $(A,B)\neq 0$. The corresponding
dispersion relation \cite{SRBS11} has the form
\begin{equation}
\label{dispersion}
{\cal E}_\pm=
\pm\sqrt{|D({\mathbf k})|^2 -\gamma^2}
\end{equation}
For $\gamma=0$ the dispersion relation is
\begin{equation}
 {\cal E}=\pm \left|D({\mathbf k})\right|
\end{equation}
and we have two bands 
of width $t+2$. There are three pairs of diabolic points,

\begin{equation}
\begin{array}{l}
 \mathbf k^{\pm,\mp}_0= (\pm \pi,\pm \arccos{
\frac{t}{2}})  \\
 \mathbf k^{\pm}_1=(0,\pm (\pi- \arccos{\frac{t}{2}}))
\end{array}  
\end{equation}
Expansion of $D(\mathbf{k})$ up 
to the first order around the DPs leads to
\begin{equation}
 D({\mathbf k}) \approx it \eta_x\mp \sqrt{4-t^2}\eta_y.
\end{equation}
Substituting the above expression in the energy dispersion given by Eqn. (\ref{dispersion}), we get the following linear relation
\begin{equation}
  {\cal E}_{\mathbf k} \approx  
\pm(t^2 \eta_x^2+(4-t^2)\eta_y^2)^{1/2}
\end{equation}
where $\eta_{x,y}= k_{x,y} - (k_{0,1})_{x,y}$.

The standard passive ($\gamma=0$) honeycomb lattice (zero strain) corresponds to $t=1$. In this case, there 
are three pairs of DPs at (see Fig.\ref{figure:fig2}-a)
\begin{equation}
\begin{array}{l}
\mathbf k^{\pm,\mp}_0=(\pm \pi,\pm \frac{\pi}{3})\\
\mathbf k^{\pm}_1=(0,\pm\frac{ 2\pi}{3})
\end{array}
\end{equation}
For $1 < t < 2$ the two pairs of DPs at $\mathbf k^{\pm,\mp}_0$ start moving toward each 
other while the pair at $\mathbf k^{\pm}_1$ moves away from one another. At $t=2$ a  {\it degeneracy} occurs i.e.  
$\mathbf k^{\pm,\mp}_0=(\pm \pi,0)$. At the same time the dispersion relation ${\cal E}$ around $\mathbf 
k^{\pm,\mp}_0$ and $\mathbf k^{\pm}_1$, is linear only in the $k_x$ direction (and quadratic in the $k_y$). 
For $t>2$ the two energy surfaces move away from each other and a gap between them is created (see Fig. \ref{figure:fig2}-b). 
Therefore for $t\geq 2$ the DPs disappeared for $\gamma=0$, and the conical diffraction is destroyed \cite{T08,note1}.

By introducing gain/loss to the system described by Eq. (\ref{dyndimer}) the resulting effective Hamiltonian 
which describes the paraxial evolution becomes non-Hermitian. In fact, for $1\leq t\leq2$, any value of $\gamma$ 
results in complex eigenvalues, i.e. the system is in the broken ${\cal PT}$-symmetry phase. The resulting dispersion 
relation, resembles the dispersion relation of relativistic particles with imaginary mass as was recently discussed
in Ref.~\cite{SRBS11}.

In the case of $t>2$ the size of the gap between the two bands can be controlled by manipulating the gain/loss 
parameter $\gamma$. In this case, there is a $\gamma$ domain, corresponding to the exact phase, for which 
the energies are real. It turns out from Eq.  (\ref{dispersion}) that the line
\begin{equation}
\gamma_{\cal PT} = t-2
\label{gammaPTline}
\end{equation}
defines the phase transition from exact to broken ${\cal PT}$-symmetry \cite{SRBS11}. The mechanism 
for this symmetry breaking is the crossing between levels, associated to the exceptional points
\begin{equation}
\begin{array}{l}
 \mathbf k^{\pm}_{EP}= (0, \pm \pi) \\
\quad \quad \text{or} \\
\mathbf k^{\pm}_{EP}= (\pm \pi,0)
\end{array}
\end{equation}
(see Fig.\ref{figure:fig2}-c) and belonging to different bands }\cite{BFKS10}: it follows from Eq. (\ref{dispersion}), that when 
$\gamma=\gamma_{\cal PT}$, the gap disappears and the two (real) levels at the "inner" band-edges become 
degenerate. Evaluation of $D(k_x,k_y)$ to second order in $(\eta_x, \eta_y)$ around the degeneracy points, 
leads to 
\begin{equation}
\label{Dexpand}
D(\eta_x,\eta_y)\approx -(\gamma_{\cal PT}+2i\eta_x+\eta^2);\quad \eta^2\equiv \eta_x^2+\eta_y^2
\end{equation}
thus resulting in the following dispersion relation
\begin{equation}
\label{disperse}
{\cal E}=\pm \sqrt{2\gamma_{\cal PT}\eta_y^2 + (2\gamma_{\cal PT}+4)\eta_x^2}.
\end{equation}
For large $\gamma_{\cal PT}$-values (i.e. $\gamma_{\cal PT}\gg 2$) one can approximate the above equation to get
\begin{equation}
 {\cal E}\approx \pm \sqrt{2\gamma_{\cal PT}}\eta.
 \label{disperseapp} 
\end{equation}

This comment will become important in the analysis of optical beam propagation discussed in the next section \ref{dynamics}.

Next, we turn to the analysis and characterization of the bi-orthogonal set of eigenvectors of our non-Hermitian system.
The target here is to identify the proximity to the exceptional point in the case of finite Hilbert spaces, where finite
size effects might play an important role in the analysis of the dynamics.
The latter do not respect the standard (Euclidian) ortho-normalization condition. Let $\langle L_{n}|$ and $|R_{n}\rangle$ 
denote the left and right eigenvectors of the non-hermitian Hamiltonian ${\cal H}$ corresponding to the eigenvalue ${\cal 
E}_n$, i.e.
\begin{equation}
\left\{
\begin{array}{l}
 \langle L_n|{\cal H}=
\langle L_n|{\cal E}_n  \\
\quad \quad \text{ and}  \\
{\cal H}|R_n\rangle= {\cal E}_n|R_n\rangle.
\end{array}  \right.
\end{equation}
 The vectors can be normalized to satisfy
 \begin{equation}
\left\langle L_{n}|R_{m}\right\rangle =\delta_{nm}
 \end{equation}
 while 
 \begin{equation}
  \sum_{n}^{\cal N}\left|R_{n} \right\rangle \left\langle L_{n}\right|=1
 \end{equation}

Above ${\cal N}$ is the dimension of the Hilbert space. 

An observable that measures the non-orthogonality of the modes, and can be used in order to identify the proximity to
the EP in the presence of finite size effects, is the so-called \emph{Petermann factor} which is defined as \cite{Petermann79}
\begin{equation}
 K_{nm}=\left\langle L_{n}| L_{m}\right\rangle \left\langle R_{m}|R_{n}\right\rangle
\end{equation}  
We have studied the mean (diagonal) Petermann factor 
\begin{equation}
 \overline{K}=\frac{1}{\cal N} \sum_{n=1}^{\cal N}K_{nn}
\end{equation}
which takes the value $1$ if the eigenfunctions of the system are orthogonal while is larger than one 
in the opposite case. At the EP, a pair of eigenvectors associated with the corresponding degenerate eigenvalues 
coalesce, leading to a  {\it collapse}  of the Hilbert space. At this point, the Petermann
factor diverges as 
\begin{equation}
  \overline{K}\sim 1/ |\gamma-\gamma_{\cal PT}|
\end{equation}
 \cite{Berry03,ZCFK10}. This indicates strong correlations between the spectrum and the eigenvectors which can 
affect drastically the dynamics as we will see later. 

\begin{figure}
   \includegraphics[width= 1\linewidth, angle=0]{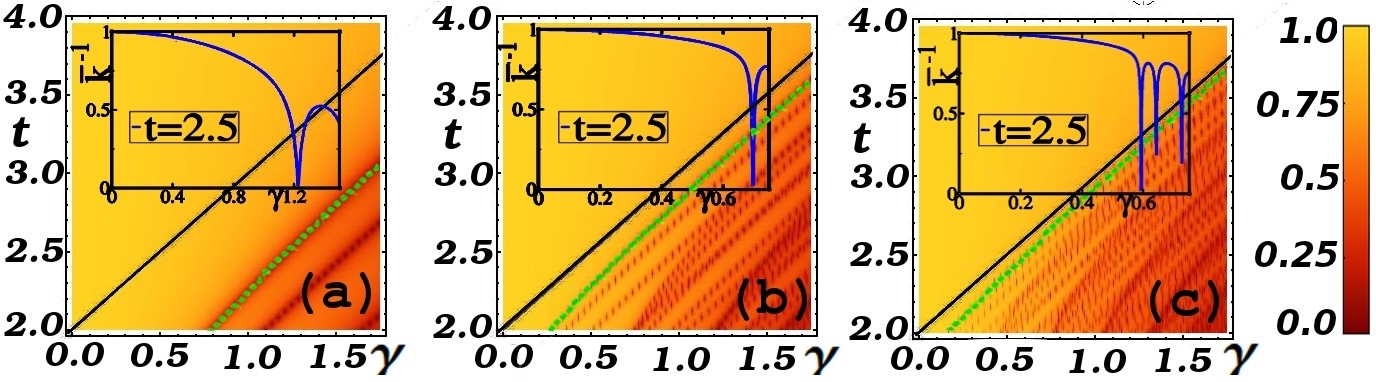}
    \caption{Inverse Petermann factor for various system sizes (a) ${\cal N}=24$; (b) ${\cal N}=168$; (c) ${\cal N}=
440$ . The asymptotic line $\gamma_{\cal PT}=t-2$ (black line) is approached by our numerical data 
(zero of $\overline{K}^{-1}$) for increasing ${\cal N}$. The green line is plotted in order to guide the eye. The 
insets show representative $\overline{K}^{-1}$'s vs. $\gamma$ for a fixed coupling $t=2.5$.
\label{figure:fig2peterman}}
\end{figure}

In Figs. \ref{figure:fig2peterman}a-c we present our calculations for $\overline{K}$ for different $(t,\gamma)$ values and 
various system sizes. We find that as ${\cal N}$ increases the divergence is approaching the line $\gamma_{\cal PT}=t-2$, which 
was derived previously (see Eq.~(\ref{gammaPTline}) above) for the case of infinite graphene.

\section{Dynamics}\label{dynamics}

Armed with the previous knowledge about the eigenmodes properties of the ${\cal PT}$-symmetric graphene, we are now ready to
study beam propagation in $\cal{PT}$-symmetric honeycomb lattices at the vicinity of the EPs. The question at hand is whether
the collapse of the Hilbert space at the EP, will affect the CD pattern, and if yes, what is the emerging dynamical picture.

\subsection{Numerical Analysis}\label{numerics}

\begin{figure}
   \includegraphics[width= 1\linewidth, angle=0]{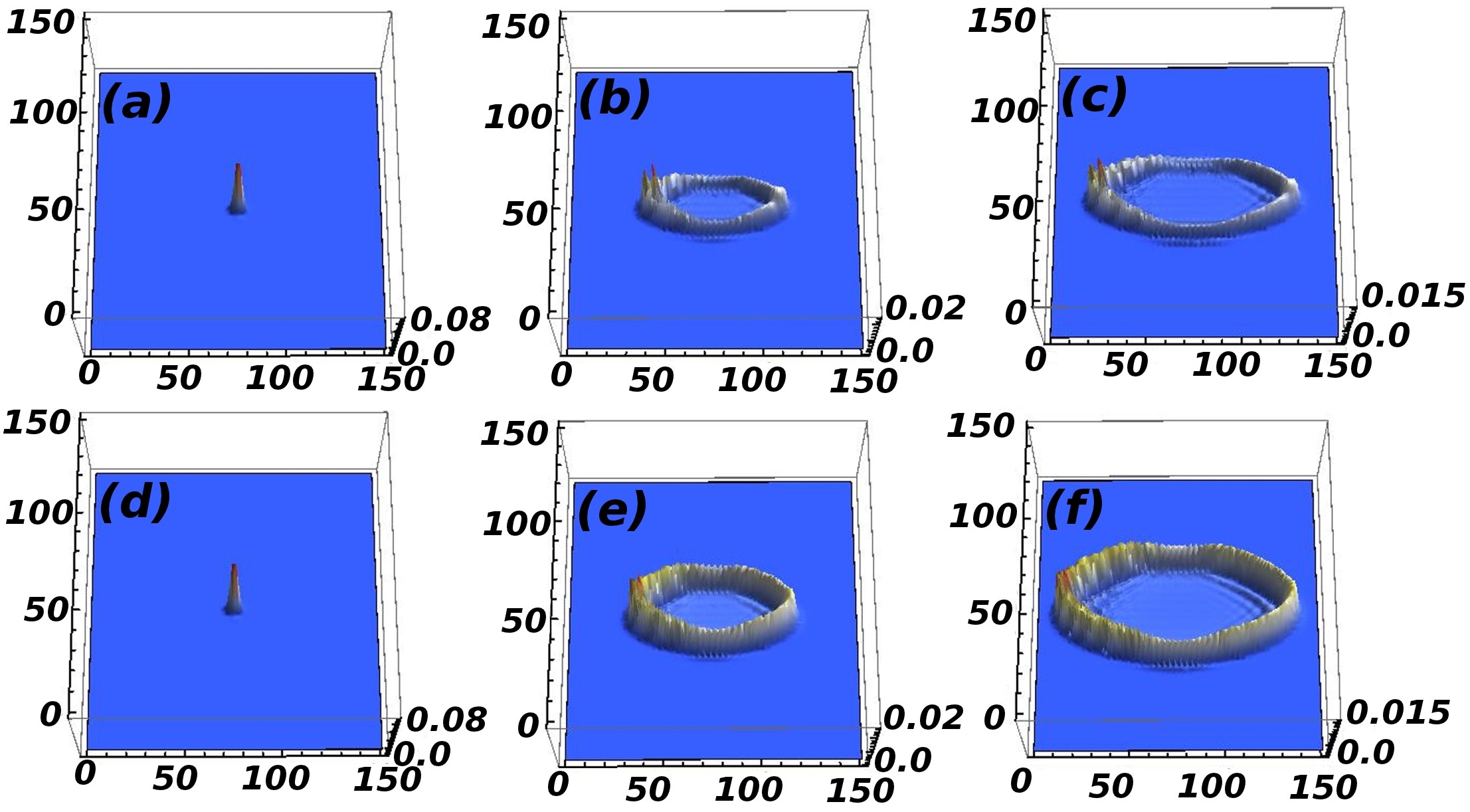}
    \caption{Propagation of a Gaussian superposition of Bloch modes associated with the vicinity of an EP at 
${\mathbf k}=(0,\pm \pi)$, in a ${\cal PT}$
-symmetric honeycomb lattice. Shown is the beam intensity at normalized propagation distances of (a,d) $z=0$; (b,e)$z=10$; 
(c,f) $z=15$, for $\gamma_{\cal PT}=1$ and $t=3$ (upper panels) and $\gamma_{\cal PT}=2$ and $t=4$ (lower panels). The 
input bell-shaped beam transforms into a ring-like structure of light of a non-varying thickness. 
\label{figure:fig3}}
\end{figure}

We first study wave propagation in the honeycomb lattice numerically (Fig. \ref{figure:fig3}), by 
launching a beam with the structure of a Bloch mode associated with the EP, multiplied by a Gaussian envelope. The Bloch 
modes at the tip can be constructed from pairs of plane waves with ${\mathbf k}$ vectors of opposite pairs of exceptional points.
 Thus, interfering two plane-waves at angles associated with opposite EP yields the phase structure of the modes from these points.
 Multiplying these waves by an envelope yields a superposition of Bloch modes in a region around these points. Figure \ref{figure:fig3} 
shows an example of the propagation of a beam constructed to excite a Gaussian superposition of Bloch modes around an EP. 
The input beam has a bell-shape structure, which, after some distance, transforms into the 
ring-like characteristic of conical diffraction \cite{note1}. From there on, the ring is propagating in the lattice by
keeping its width constant while its radius is increasing linearly with distance. The invariance of the ring thickness 
and structure 
manifests a (quasi-) linear dispersion relation above and below the EP (see Fig. 2); hence, the diffraction coefficient 
for wave packets constructed from Bloch modes in that region is zero (infinite effective mass). This is especially 
interesting because the ring itself is a manifestation of the dispersion properties at the EP itself, where the diffraction 
coefficient is infinite (zero-effective mass). As a result, the ring forms a light cone in the lattice. The appearance of 
CD in the case of ${\cal PT}$-lattices where the eigenvectors are non-orthogonal and coalesce at the EP singularity, provides 
a clear indication that the phenomenon is insensitive to the eigenmodes structure and it depends only on the properties
of the dispersion relation.

The ${\cal PT}$-symmetric conical diffraction shows some unique characteristics with respect to the CD found in the case 
of beam propagation around DPs for passive honeycomb lattices (i.e. $\gamma=0$). A profound difference is associated with 
the fact that now, the transverse speed of the cone is increased \cite{SRBS11} and in fact it can be controlled by the magnitude of the 
gain/loss parameter at the symmetry breaking point $\gamma_{\cal PT}$. This is shown in Fig. \ref{figure:fig3} 
where we compare the spreading of a CD for two different $\gamma_{\cal PT}$-values.

\begin{figure}
   \includegraphics[width= 0.81 \linewidth, angle=0]{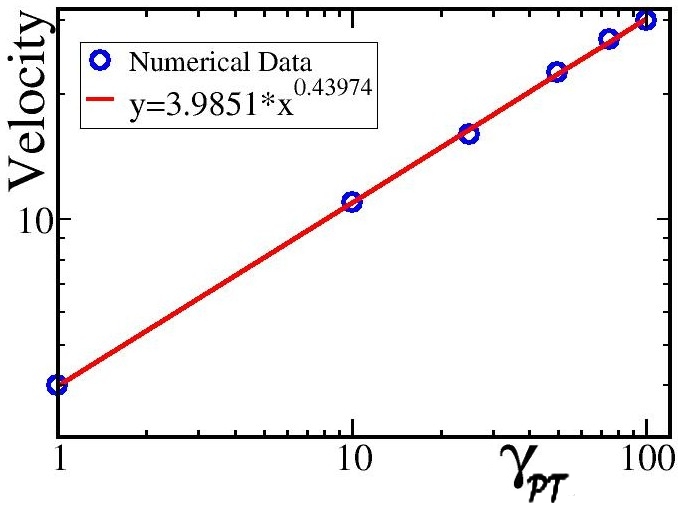}
    \caption{Transverse velocity of the spreading ring versus the $\gamma_{\cal{PT}}$.
     Numerical simulation approve theoretical prediction for the transverse speed of 
     the CD, which is $v_g \propto \sqrt{\gamma_{\cal PT}}$. 
\label{figure:fig4}}
\end{figure}

In Figure \ref{figure:fig4}, we report in a double logarithmic plot, our numerical measurements for the transverse velocity 
of the spreading ring for various $\gamma_{\cal{PT}}$ values. The best linear fitting to the numerical data shows that transverse 
speed of the cone is proportional to $ \sqrt{\gamma_{\cal PT}}$. This behavior, can be understood qualitatively by realizing that 
the group velocity near the EP's is 
\begin{equation}
 v_g=\partial {\cal E}/\partial \mathbf k\sim \sqrt{2\gamma_{\cal PT}} 
\end{equation}
(see Eq. (\ref{disperse}) and Eq.(\ref{disperseapp})).
At the same time we find that the resulting ${\cal PT}$-cone is brighter with respect to the one found in passive 
lattices \cite{P07} (i.e. the field intensity of the conical wavefront is larger) .


\subsection{Theoretical Considerations}\label{theory}

It is possible to gain valuable insight into the features of ${\cal PT}$-conical diffraction by considering the field 
evolution in the momentum space. We consider for simplicity, an initial distribution $({\tilde a}_{\mathbf k}(0), 
{\tilde b}_{\mathbf k}(0))^T$ that is symmetric around the EP while it decays exponentially away from it. 
Specifically we assume 
\begin{equation}
 ({\tilde a}_{\mathbf k}(0), {\tilde b}_{\mathbf k}(0))^T=e^{-g \sqrt{\eta_{x}^2 + \eta_{y}^2}} 
(1,0)^T. 
\end{equation}
 Next, we calculate the evolution matrix $\mathbf{U}\equiv e^{-i \mathbf{H_k}z}$ where $\mathbf{H_k}$ is given 
by Eq.~(\ref{hamilt}). After a straightforward algebra and using the fact that
\begin{equation}
\mathbf{H_k}^2= {\cal E}^2 \times \mathbf{1},
\end{equation}
where $\mathbf{1}$ is the unity matrix, we get 
\begin{equation}
\label{evolve}
\mathbf{U}=\cos(z|{\cal E}|)\mathbf{1}-i(\sin(z|{\cal E}|)/|{\cal E}|)
\mathbf{H_k}
\end{equation}
Equation~(\ref{evolve}) is the starting point of our analysis. Substituting Eqs. (\ref{hamilt},\ref{Dexpand},
\ref{disperse}), we find that the evolving amplitude of the field $(a_{n,m}, b_{n,m})$ is
\begin{eqnarray}
\label{final}
a_{n,m}(z)&\approx& \sum_{l=1,2} 
\frac{(-1)^{l}i\left[z-\gamma_{\cal PT} \phi(n,m,z,g)\right]+ g} {\phi(n,m,z,g)^{3/2}} \\
b_{n,m}(z)&\approx& \sum_{l=1,2} 
\frac{(-1)^{l+1}\left(\gamma_{\cal PT} \phi(n,m,z,g) + \frac{n}{\sqrt{2\gamma_{\cal PT}+4}}\right)} {\phi(n,m,z,g)^{3/2}} \nonumber 
\end{eqnarray}
where 
\begin{equation}
 \phi(n,m,z,g)=\left[g+(-1)^{l}iz\right]^2+n^2/(2\gamma_{\cal PT}+4)+m^2/(2\gamma_{\cal PT}).
\end{equation}
 Although our simplified calculations are  not able to capture all  the features of the propagating cone, 
 the above expression encompasses the main characteristics of the conical diffraction 
that we have observed in our numerical simulations. At $z= 0$, Eq. (\ref{final}) resembles a Lorentzian, which slowly 
transforms into a ring of light, whose radius expands linearly with $z$ with velocity $\sqrt{2\gamma_{\cal PT}}$, while 
its thickness  remains unchanged. At the same time the field intensity on the ring in the case of ${\cal PT}$-symmetric lattices is 
brighter than the one corresponding to passive honeycomb lattices (i.e. $1/z^2$ vs $1/z^4$ behavior respectively). 

\section{Summary and Concluding Remarks}\label{conclusions}

We studied numerically and analytically the propagation of waves in ${\cal PT}$-honeycomb photonic lattices, demonstrating 
the existence of conical diffraction arising solely from the presence of a spontaneous ${\cal PT}$-symmetry breaking phase 
transition point. In spite the fact that the eigenvectors are non-orthogonal and there is a collapse of the Hilbert
space at the EP, the emerging cone, is brighter and moves faster than the corresponding one of the passive structure. 
Although, the realization of such photonic structures is currently a challenging task, active electronic circuits, like
the one proposed in Ref.~\cite{SLZEK11}, can be proven useful alternatives that might allow us to investigate experimentally,
wave propagation in extended ${\cal PT}$-symmetric lattices.

These findings raise several intriguing questions. For example, how does nonlinearity affect ${\cal PT}$-symmetric conical 
diffraction? What is the effect of disorder \cite{BFKS09}? Is this behavior generic for any system at the spontaneously 
${\cal PT}$-symmetry breaking point? These intriguing questions are universal, and relate to any field in which waves 
can propagate in a periodic potential. It is expected that  in active metamaterials  such phenomena will be present and 
they may have specific technological importance.

\begin{acknowledgments}
{\it Acknowledgments --}
(TK) and (HR) acknowledge support by a AFOSR No. FA 9550-10-1-0433 grant and (DNC) by a AFOSR No. FA 9550-10-1-0561 grant. 
VK's  work was supported  via AFOSR LRIR 09RY04COR.
\end{acknowledgments}




\begin{thebibliography}{99}

\bibitem{BJ07} M. V. Berry and M. R. Jeffrey, Prog. Opt. {\bf 50}, 13 (2007); M. V. Berry, M. R. Jeffrey, and J. 
G. Lunney, Proc. R. Soc.  London, Ser. A {\bf 462}, 1629 (2006).

\bibitem{IN06} R. A. Indik and A. C. Newell, Opt. Express {\bf 14}, 10614 (2006).

\bibitem{H37} W. R. Hamilton, Trans. R. Irish Acad. {\bf 17}, 1 (1837).

\bibitem{L37} H. Lloyd, Trans. R. Irish Acad. {\bf 17}, 145 (1837).

\bibitem{P07} O. Peleg et al., Phys. Rev. Lett. {\bf 98}, 103901 (2007)

\bibitem{T08} O. Bahat-Treidel et al., Opt. Lett. {\bf 33}, 2251 (2008)

\bibitem{ANZ09}M. J. Ablowitz, S. D. Nixon,  Y. Zhu, Phys. Rev. A {\bf 79}, 053830 (2009); M. J. Ablowitz, Y. Zhu
Phys. Rev. A {\bf 82}, 013840 (2010). 

\bibitem{E06} R. I. Egorov {\it et al.}, Opt. Lett. {\bf 31}, 2048 (2006).

\bibitem{BPGSSP10} O. Bahat-Treidel, O. Peleg, M. Grobman, N. Shapira, M. Segev, and T. Pereg-Barnea, Phys. Rev. Lett. 
{\bf 104}, 063901 (2010).

\bibitem{SRBS11} A. Szameit {\it et al.}, Phys. Rev. A {\bf 84}, 021806 (2011). 

\bibitem{MGCM08a} K. G. Makris {\it et. al}, Phys. Rev. Lett. {\bf 100}, 103904 (2008).

\bibitem{K10} T. Kottos, Nature Physics {\bf 6}, 166 (2010).

\bibitem{BB98} C. M. Bender, S. Boettcher, Phys. Rev. Lett. {\bf 80}, 5243 (1998);
C. M. Bender, D. C. Brody, H. F. Jones, Phys. Rev. Lett. {\bf 89}, 270401 (2002).

\bibitem{BBDJ02} M. Znojil, Phys. Lett. A {\bf 285}, 7 (2001); H. F. Jones, J. Phys. A  {\bf 42}, 135303 (2009).

\bibitem{B07} C. M. Bender, Rep. Prog. Phys. {\bf 70}, 947 (2007);
C. M. Bender {\it et. al}, Phys. Rev. Lett. {\bf 98}, 040403 (2007).


\bibitem{RMGCSK10} C. E. Ruter {\it et. al}, Nat. Phys. {\bf 6}, 192 (2010).

\bibitem{GSDMRASC09} A. Guo, {\it et. al.}, Phys. Rev. Lett. {\bf 103}, 093902 (2009).

\bibitem{SLZEK11} J. Schindler, {\it et. al}, Phys. Rev. A {\bf 84}, 040101(R) (2011).

\bibitem{FAHXLCFS11}L. Feng, M. Ayache, J. Huang, Y.-L. Xu, M.-H. Lu, Y.-F. Chen, Y. Fainman, A. Scherer,
Science {\bf 333}, 729  (2011)

\bibitem{ZCFK10} M. C. Zheng {\it et. al},  Phys. Rev. A {\bf 82}, 010103 (2010).

\bibitem{JG11} H. Jones and E. M. Gr\"afe, Phys. Rev. A (2011).

\bibitem{L09a}S. Longhi, Phys. Rev. Lett. {\bf 103}, 123601 (2009).

\bibitem{LRKCC11} Z. Lin, {\it et. al}, Phys. Rev. Lett {\bf 106}, 213901 (2011)

\bibitem{RKGC10} H. Ramezani {\it et. al}, Phys. Rev. A {\bf 82}, 043803 (2010)

\bibitem{L10b} S. Longhi, Phys. Rev. A {\bf 82}, 031801 (2010);
Y. D. Chong, L. Ge, A. D. Stone, Phys. Rev. Lett. {\bf 106}, 093902 (2011)

\bibitem{FMM} F. Nazari, M. Nazari, M. K. Moravvej-Farshi, Opt. Lett. {\bf 36}, 4368-4370 (2011).

\bibitem{SXK10} A. A. Sukhorukov, Z. Xu, Y. S. Kivshar, Phys. Rev. A {\bf 82}, 043818 (2010).

\bibitem{Mussli}Z. H. Musslimani {\it et. al}, Phys. Rev. Lett. {\bf 100},
030402 (2008).

\bibitem{M02a} A. Mostafazadeh,
J. Math. Phys. {\bf 43}, 205 (2002)

\bibitem{M02b} A. Mostafazadeh,
J. Math. Phys. {\bf 43}, 3944 (2002)

\bibitem{M08} A. Mostafazadeh,
J. Phys. A {\bf 41}, 055304 (2008)

\bibitem{BBM02} C. M. Bender, M. V. Berry, A. Mandilara,
J. Phys. A {\bf 35}, L467 (2002).

\bibitem{BFKS10} O. Bendix {\it et al.}, J. Phys. A: Math. Theor. {\bf 43}, 265305 (2010).

\bibitem{note1} Here we use the term "conical" diffraction in a rather loose fashion. Specifically, 
when the lattice is deformed i.e. $t\neq 1$, the CD pattern becomes elliptic \cite{T08}.

\bibitem{Petermann79} K. Petermann, IEEE J. Quantum Electron. {\bf 15}, 566 (1979); A. E. Siegman, Phys.
Rev. A {\bf 39}, 1264 (1989).

\bibitem{Berry03}  M. V. Berry, Journal of Modern Optics {\bf 50}, 63 (2003);
S.-Y. Lee {\it et al.}, Phys. Rev. A {\bf 78}, 015805 (2008).

\bibitem{BFKS09} O. Bendix, {\it et. al}, Phys. Rev. Lett. {\bf 103}, 030402 (2009); 
C. T. West, T. Kottos, T. Prosen, {\it ibid.} {\bf 104}, 054102 (2010).

\end{thebibliography}
\end{document}